**Ion-Implanted Silicon Nanoregions Enable Ultra-Low-Loss Trimming of Cladded Photonic Integrated Circuits**

*Zhongyu Tang, Shabnam Taheriniya, Seongmin Jo, Xinyu Ma, Akhil Varri, Anna P. Ovvyan, Vincent Spreter, Liam McRae, Xiansong Meng, Wolfram H. P. Pernice*, Rongyang Xu**

Z. Tang, S. Taheriniya, S. Jo, X. Ma, A. P. Ovvyan, V. Spreter, L. McRae, X. Meng, W. H. P. Pernice, R. Xu
Kirchhoff-Institute for Physics, Heidelberg University, Heidelberg, Germany
E-mail: wolfram.pernice@kip.uni-heidelberg.de, rongyang.xu@kip.uni-heidelberg.de

S. Jo, A. Varri
Institute of Physics, University of Münster, Münster, Germany

Photonic integrated circuits (PICs) have emerged as a key platform for information processing, including optical communication and computing. As PIC complexity increases, fabrication-induced response deviations accumulate, making post-fabrication trimming critical for unlocking their full potential. Here, we demonstrate that silicon ion implantation enables scalable, ultra-low-loss trimming of cladded PICs by locally forming a high-index silicon-rich region. Structural characterization confirms that the silicon-implanted nanoregion is confined to the cladding without observable damage to the underlying waveguide. The implantation-induced excess loss is below 0.001 dB per π phase shift, while the optical response remains stable over a four-month observation period. Automated trimming is demonstrated on a photonic crossbar array, reducing the average channel output variation from 78.5% to 5.9%. These results establish a practical route towards automated, ultra-low-loss post-fabrication trimming for large-scale cladded PICs.

## 1 | Introduction

Driven by their low latency[1,2], high energy efficiency[3,4], and inherent multiplexing capability[5–11], photonic integrated circuits (PICs) are becoming increasingly prevalent across fields ranging from telecommunications[12–15] and quantum information processing[16–18] to next-generation artificial intelligence hardware[5,19]. However, realizing the full potential of PICs remains challenging because their optical responses are highly sensitive to nanometer-scale fabrication variations[20], causing response deviations to accumulate in large-scale PICs and degrade overall system functionality[21,22]. Post-fabrication trimming is widely employed to compensate for these variations and ensure reliable performance.

Post-fabrication trimming techniques are broadly categorized into volatile and non-volatile approaches. Volatile approaches primarily rely on the thermo-optic effect, requiring continuous power to modify the refractive index[23–25]. Non-volatile approaches offer permanent corrections without continuous power consumption by locally modifying the optical properties of waveguides. For example, phase change materials (PCMs) deposited on waveguides enable reversible tuning through phase transitions[5,26]. Ge ion implantation enables refractive index modification of silicon waveguides through silicon amorphization[27]. UV laser irradiation has also been used to alter the microstructure of silicon nitride waveguides[28]. These approaches have enabled low-loss, non-volatile tuning and significantly advanced the development of post-fabrication trimming technologies. Top cladding is widely employed in practical PICs to reduce propagation loss, enhance environmental robustness, and ensure long-term device reliability[29,30]. Extending tuning to cladded PICs while maintaining high spatial resolution and ultra-low excess loss remains an important challenge.

In this work, we demonstrate a non-volatile, cross-platform post-fabrication trimming technique for cladded PICs using automated focused ion beam (FIB) silicon ion implantation. We validate the technique using representative photonic components, including directional couplers and ring resonators. It provides a broad tuning range with an excess loss below 0.001 dB per $\pi$ phase shift while remaining stable under ambient conditions. The automated trimming approach is scalable, extending from individual devices to photonic circuits, as demonstrated on a photonic crossbar array.

## 2 | Results

### 2.1 | Mechanism

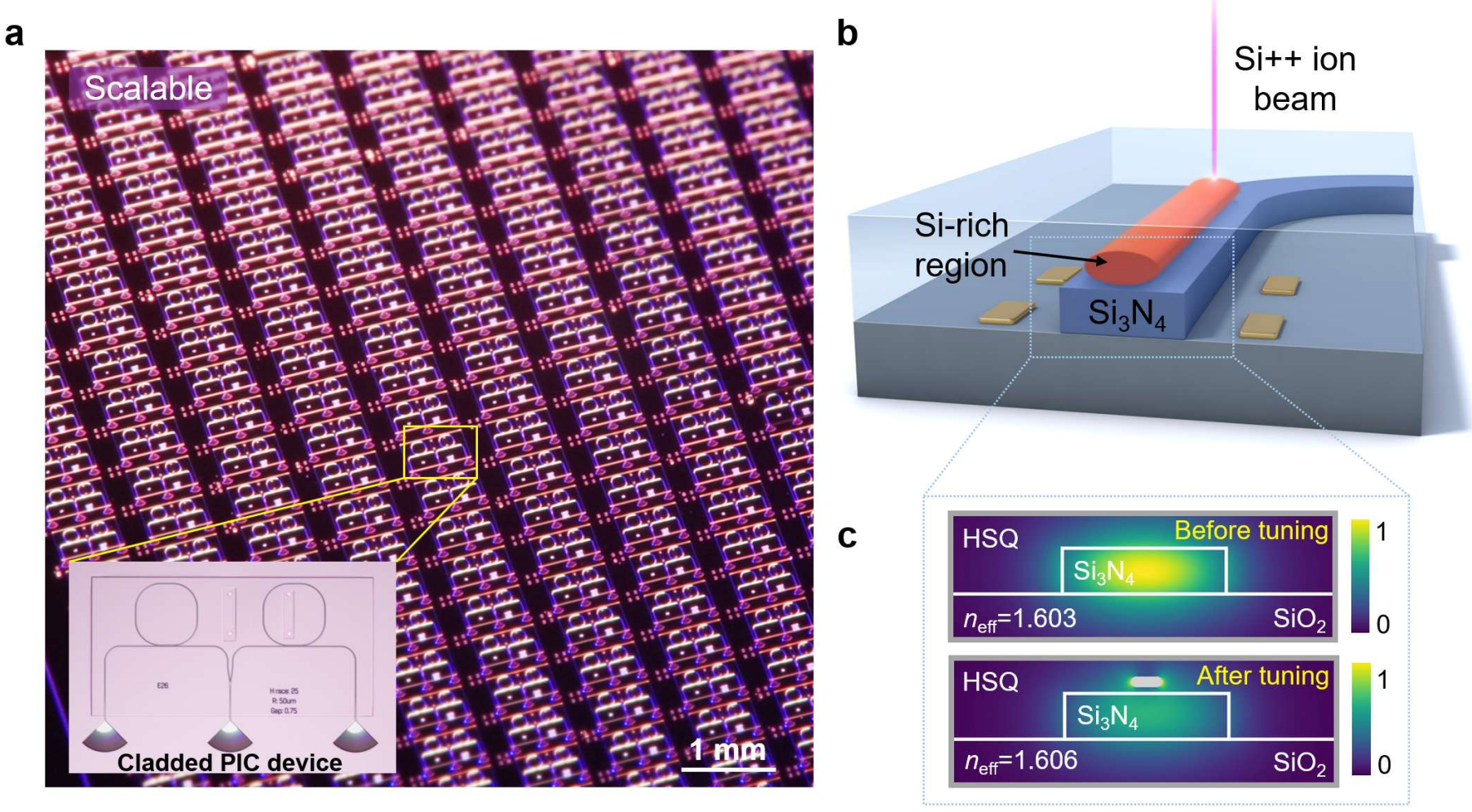

**FIGURE 1 |** Schematic and mechanism of silicon ion implantation in cladded photonic integrated circuits. (a) Photograph of a fabricated PIC chip containing cladded device arrays. (b) Schematic of the device cross-section where ion implantation is performed. The Si++ ion beam forms a Si-rich region confined to the cladding, without reaching the waveguide underneath. (c) Simulated mode profile without (top) and with (bottom) the addition of a Si-rich region above the waveguide.

The ion implantation process on a photonic chip is illustrated in Figures 1a,b. Hydrogen silsesquioxane (HSQ), whose optical properties after curing closely resemble those of $SiO_2$, is used as the top cladding. Si++ ions from a liquid metal alloy ion source (LMAIS) are accelerated to 70 keV and implanted into the HSQ top cladding, forming a Si-rich amorphous nanoregion above the waveguide. The evanescent field of the guided mode overlaps with the Si-rich amorphous region, thereby tuning the effective refractive index $n_{eff}$ of the guided mode without directly modifying the waveguide core. Automated alignment enables sequential trimming of multiple devices using predefined implantation patterns. Figure 1c shows the simulated mode profiles of a cladded silicon nitride ($Si_3N_4$) waveguide before and after the introduction of a 70 nm-wide silicon region above the waveguide. Due to the higher refractive index of silicon compared with $Si_3N_4$, the $n_{eff}$ of the guided mode increases from 1.603 to 1.606. The spatial distribution of implanted silicon within the cladding was simulated (Supporting Information Section S1). At an acceleration voltage of 35 kV, the center of the silicon-implanted region is located 100 nm below the cladding surface. This implantation profile suggests that the approach is suitable for top-cladding thicknesses of less than 1 µm, while maintaining sufficient overlap with the evanescent field of the guided mode.

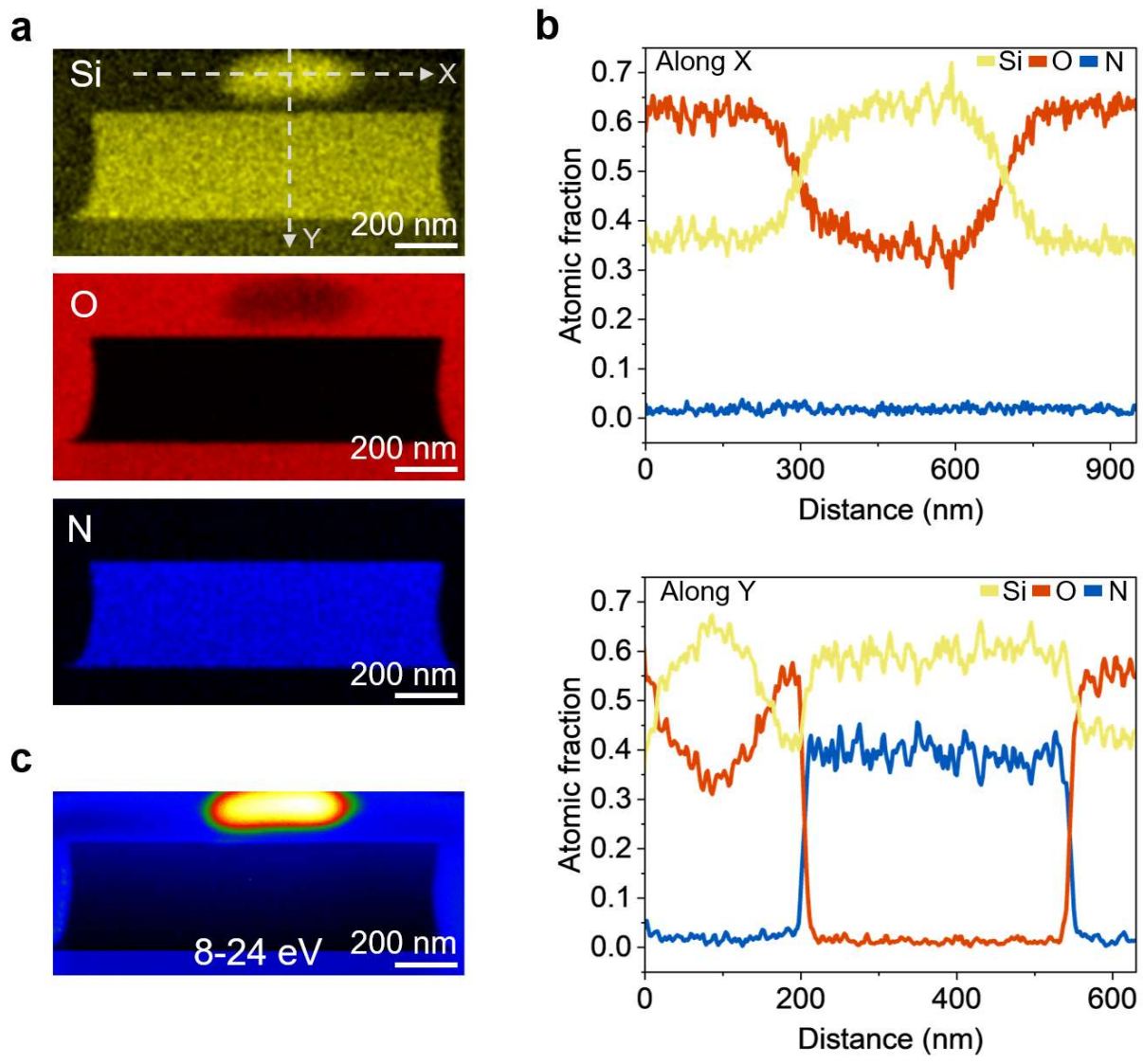

**FIGURE 2 |** Transmission electron microscope study of the silicon implanted device. (a) STEM-EDS elemental maps of a cladded $Si_3N_4$ device cross-section after implantation, showing the Si-rich, O-depleted implanted area. The HSQ cladding above the $Si_3N_4$ waveguide is approximately 150 nm thick. (b) EELS line scan profiles along the dashed arrows in (a). (c) Plasmon map of the cross-sectional region.

To confirm the effect of the implantation on local material composition and waveguide layer stack, energy-dispersive X-ray spectroscopy (EDS) and electron energy loss spectroscopy (EELS) analyses were performed using a scanning transmission electron microscope (STEM). Figure 2a shows the EDS elemental maps of the implanted waveguide cross-section for Si, O, and N, respectively. A Si-rich, O-deficient region is visible above the waveguide center, which corresponds well with the simulated Si implantation profile (Supporting Information Section S1), confirming the intended implantation depth. This compositional change indicates the accumulation of implanted Si accompanied by local oxygen depletion, suggesting that implantation modifies the local chemical environment within the HSQ cladding. Figure 2b presents the line scan profiles corresponding to the atomic concentration of the 3 elements along a horizontal (X) and a vertical (Y) line. Both plots show a local increase in Si concentration in the implantation region only, and no significant elemental changes are found outside the Si-rich cluster. The linescan profiles demonstrate that the implantation-induced compositional modifications remain spatially confined to the targeted region.

To independently assess the spatial extent of the implanted region, low-loss EELS mapping was performed. Figure 2c displays the plasmon map corresponding to the integrated signal over the 8–24 eV energy window associated with the Si-implanted region. The map shows that the plasmonic signature characteristic of this region is confined within the intended implantation area and is not detectably present in the waveguide. The Si-implanted region and the waveguide exhibit distinct low-loss EELS spectral signatures in this energy range, indicating different local electronic structures and compositions. Within the detection limits of the measurement, this indicates that the Si implantation does not measurably diffuse beyond the designed implantation bounds. These findings suggest that the implantation is confined to the local cladding area only, and there is minimal effect on the integrity of the waveguide underneath. The non-intrusive nature of the technique is demonstrated, and the technique is expected to be applicable to a wide range of material platforms without the need for individual adjustments.

To evaluate surface modification induced by silicon ion implantation, atomic force microscopy (AFM) measurements were performed. The measurements reveal a ~45 nm deep trench on the cladding surface, indicating partial surface milling by the ion beam during implantation (Supporting Information Fig. S1). Because the trench is confined to the surface of the low-index HSQ cladding, its overlap with the evanescent field of the guided mode is minimal, resulting in a negligible impact on device performance.

### 2.2 | Tuning of coupling-based photonic components

We first demonstrate tuning of evanescently coupled components, where the effective index modulation changes the coupling conditions between waveguides. A directional coupler (DC) is used as a representative device to validate this mechanism. Figure 3a shows the optical microscope image of a fabricated DC, with an initial coupling ratio (CR) of approximately 55%. Light is coupled into the device through the left grating coupler and exits through the two grating couplers on the right, while the transparent box outlines the HSQ cladding region. Implantation was performed at the coupling region in two configurations, shown as enlarged images in the insert of Figure 3a. By keeping the width constant and changing length, a wide range of tuning was achieved for both directions. Configuration 1 corresponds to implantation centered above the bus waveguide. In this case, the phase-matching condition in the implanted region is no longer satisfied, which weakens the local coupling and ultimately reduces the overall CR of the DC. Figure 3b shows the tuned CR values as a function of implantation length over 0-8 μm for a coupling length of 16.5 μm, demonstrating a continuous reduction in CR with increasing implantation length. In configuration 2, implantation in the gap region between the two waveguides enhances the coupling strength, thereby shortening the crossover length and increasing the CR. As shown in Figure 3c, the tuned CR values increase with increasing implantation length. These two configurations provide complementary tuning behaviors, enabling bi-directional tuning. The implantation results in average tuning slopes of approximately -3.6% and +1.5% CR per μm, respectively, with an implantation region of 0.3-μm width and 8-μm length. The method is also applicable to DCs with different initial CRs, including 25% and 33% (Supporting Information Section S2).

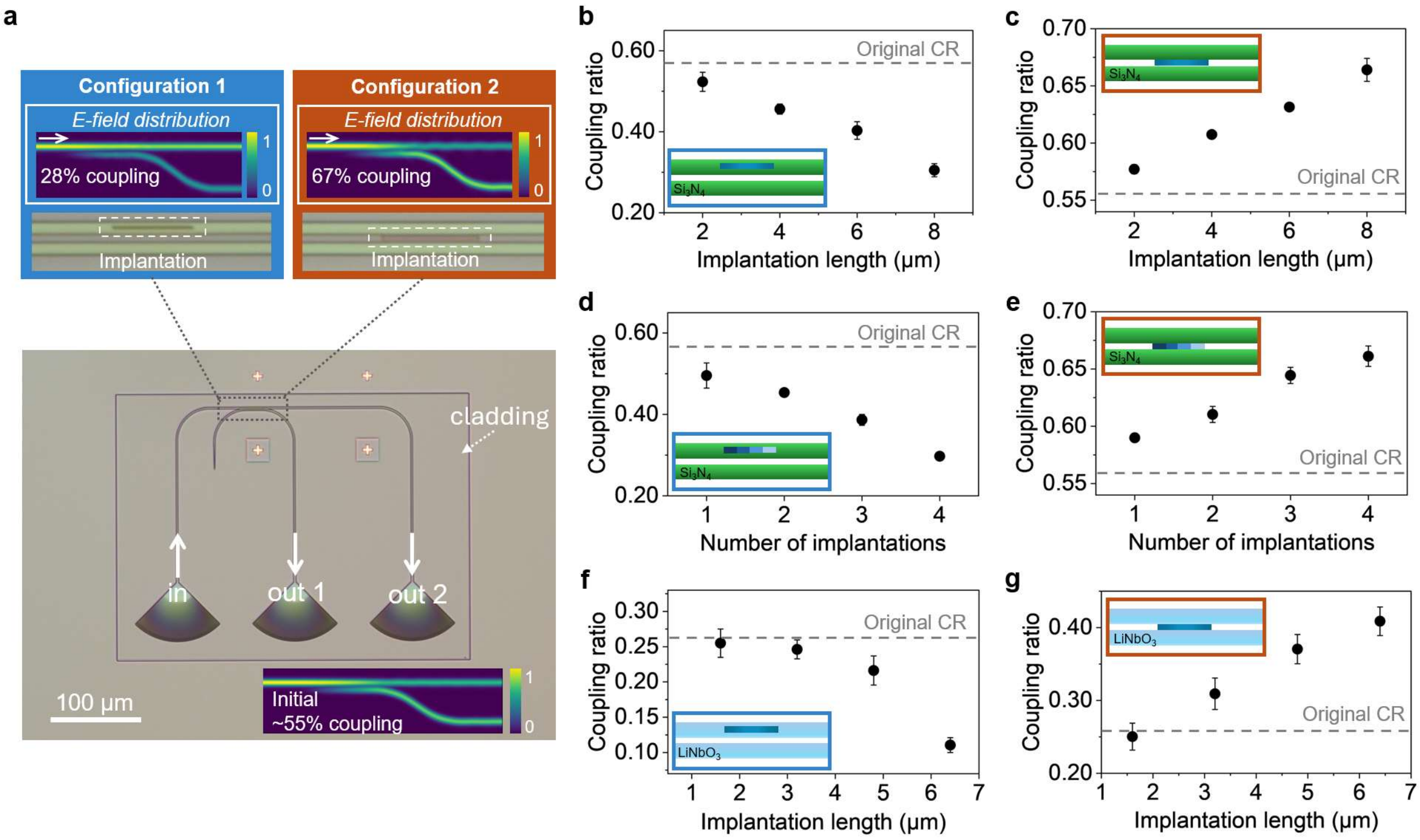

**FIGURE 3 |** Coupling ratio tuning of directional couplers by silicon ion implantation. (a) Optical microscope image of a cladded $Si_3N_4$ DC with an initial CR of ~55%. Inserts show the magnified images of the implantation region for two configurations, alongside simulated electric field distributions for an 8 μm implantation length. Configuration 1 reduces the CR, whereas configuration 2 increases it. (b,c) CR as a function of single-implantation length for configurations 1 and 2 on the $Si_3N_4$ platform, respectively. (d,e) CR after successive 2-μm implantations for the two configurations on the $Si_3N_4$ platform. The colored segments indicate the successive implantation steps. (f,g) CR as a function of implantation length for the two configurations on the $LiNbO_3$ platform.

Notably, the same device can be repeatedly tuned, enabling precise stepwise control of its optical response. Four adjacent 2 μm implantations were sequentially applied to 55% DCs, with the CR measured after each step for both configurations (Figure 3d,e). Sequential implantations produce CR variations comparable to those achieved in single-step processes with the same total implantation length.

To validate the versatility of the proposed approach, we further demonstrate the implantation-based tuning on a thin-film lithium niobate ($LiNbO_3$) platform. Figures 3f,g show the CR as a function of implantation length for both configurations, with an initial value of approximately 25%. Due to the larger sidewall angle of the lithium niobate waveguide, the implanted region interacts more strongly with the guided mode, reducing the implantation length required to achieve the same level of tuning. Despite the shorter implantation length, the tuning trend is consistent with that observed on the $Si_3N_4$ platform.

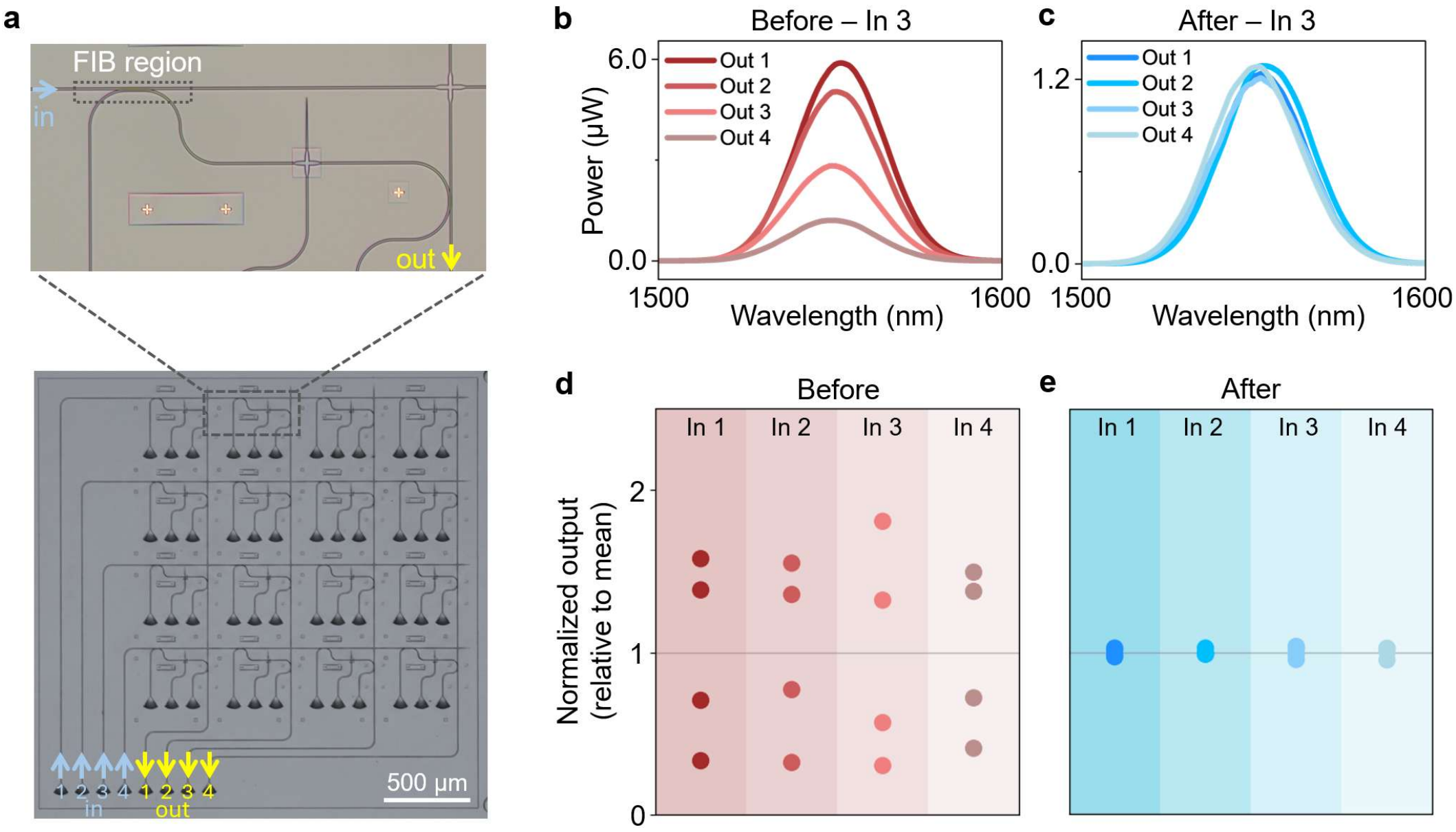

**FIGURE 4 |** Automated post-fabrication trimming of a photonic crossbar array using silicon ion implantation. (a) Optical microscope image of a 4×4 PCA with a magnified view of one cell. (b,c) Transmission spectra for input 3 before and after automated trimming. (d,e) Normalized output power distribution for each input channel before and after automated trimming.

Automated silicon ion implantation was performed on a photonic crossbar array (PCA), validating an automated trimming workflow for more complex photonic circuits. The PCA used here comprises 4 inputs, 4 outputs, and 16 cells, with silicon ion implantation performed on the horizontal DC in each cell, as shown in Figure 4a. The four DCs of each input channel are designed to achieve identical output powers. However, fabrication variations accumulate across the cascaded DCs, resulting in significant output variations. For example, Figure 4b shows the transmission spectra of the 4 outputs for input channel 3, where the maximum output power is approximately 6 times the minimum output power. A similar behavior is observed in the other input channels. Figure 4d summarizes the output power distributions normalized to the mean output power of the corresponding input channel. To compensate for these variations, the ion beam is automatically aligned to each target DC, followed by silicon ion implantation with its specified implantation length (Supporting Information Fig. S3 shows complete transmission spectra before and after trimming). Figures 4c,e show the transmission spectra of the 4 outputs in input channel 3 and the normalized output distribution for all input channels after trimming, respectively. The average output variation within each input channel is reduced from 78.5% to 5.9%. Depending on the implantation conditions, each trimming step typically requires several tens of seconds to a few minutes, making the automated workflow practical for scalable post-fabrication trimming in research and prototyping environments.

### 2.3 | Tuning of resonant photonic components

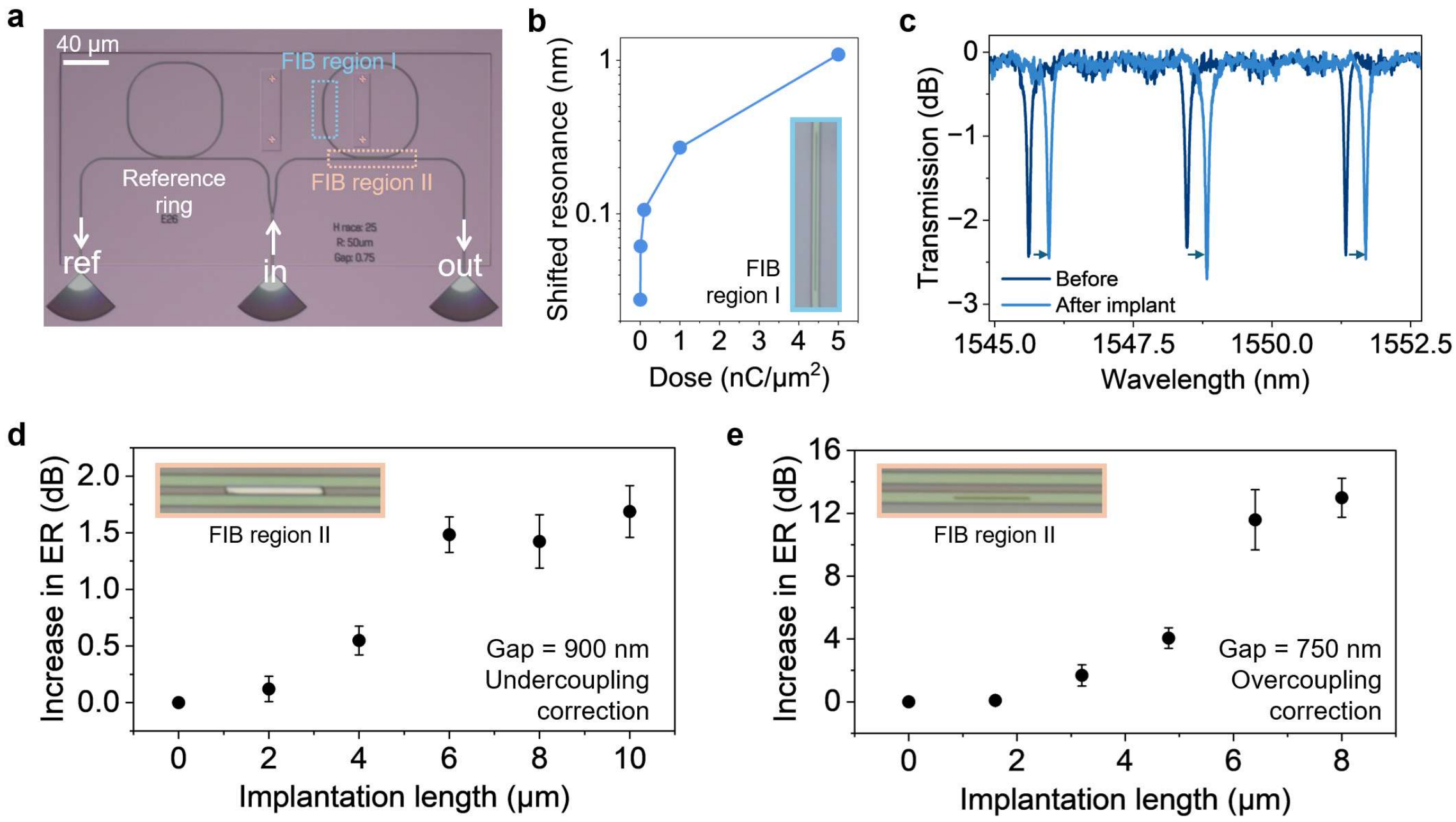

**FIGURE 5 |** Post-fabrication trimming of ring resonators. (a) Optical microscope image of a cladded microring resonator with two FIB implantation regions. (b) Resonance wavelength shift as a function of FIB dose in region I with a fixed implantation width of 100 nm and length of 25 μm. (c) Transmission spectra of the ring resonator before and after silicon ion implantation at a dose of 1 nC/μm$^2$. (d,e) Increase in ER as a function of implantation length during undercoupling and overcoupling correction at FIB region II, respectively.

Microring resonators provide a stringent test of the proposed approach because of their high sensitivity to fabrication-induced variations in both the optical phase and the coupling strength. Position-dependent implantation enables independent control of the resonance wavelength ($\lambda_{res}$) and extinction ratio (ER) in a microring resonator. Implantation on the microring resonator (region I in Figure 5a) locally modifies the effective index and optical path length, thereby tuning the $\lambda_{res}$. For a 100 nm × 25 μm fixed implantation geometry, the resonance shift increases with implantation dose (Figure 5b). At a dose of 1 nC/μm$^2$, a pronounced resonance shift is clearly observed in the transmission spectra (Figure 5c).

Selective implantation within the coupling region (region II in Figure 5a) can strengthen or weaken the coupling, allowing the coupling condition to be tuned toward critical coupling. Our devices are designed to achieve critical coupling at a gap of 0.8 μm. To demonstrate the tuning capability, devices with gap sizes of 0.75 μm and 0.9 μm were fabricated. For devices with a 0.9 μm gap, implantation within the gap strengthens the coupling, driving the undercoupled devices toward critical coupling. As a result, the ER increases with the implantation length, reaching an improvement of 1.6 dB after a 10 μm implantation (Figure 5d). For devices with a 0.75 μm gap, implantation on the bus waveguide weakens the evanescent coupling between the microring and the bus waveguide, driving the overcoupled devices toward critical coupling. As a result, the ER increases with implantation length, reaching an improvement of up to 12 dB after an 8 μm implantation (Figure 5e).

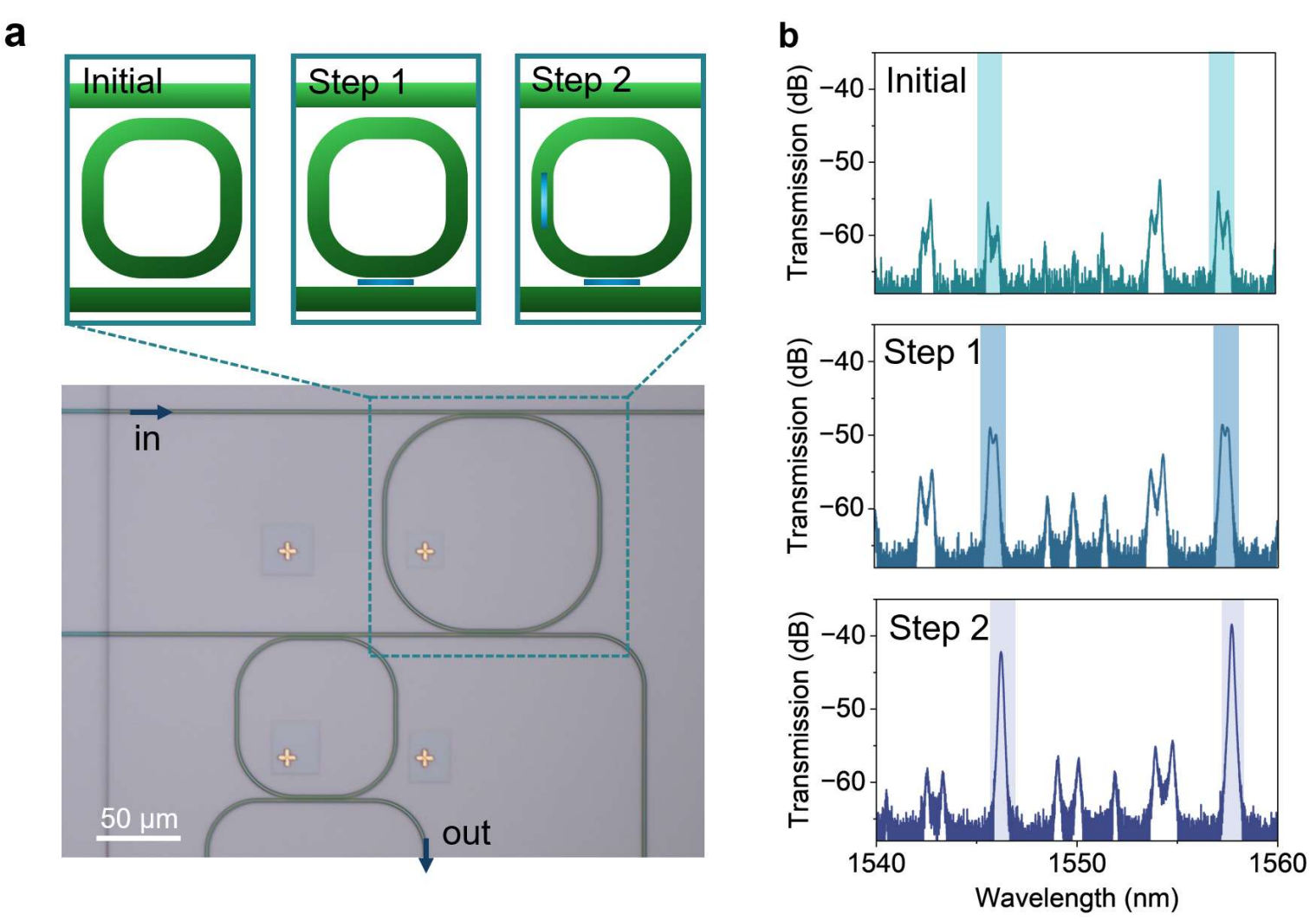

**FIGURE 6 |** Precise trimming of a Vernier filter using silicon ion implantation. (a) Optical microscope image of the fabricated Vernier filter. Inserts show the implantation locations for step 1 (ER enhancement) and step 2 (resonance shifting). (b) Transmission spectra of the device after each trimming step. Step 1, consisting of implantation in the coupling region, increases the ER by ~5 dB, while step 2 provides an additional ~ 10 dB ER enhancement through resonance alignment.

We further demonstrate that the proposed approach can be used to precisely align and enhance the resonance of a Vernier filter comprising two rings of different sizes. Figure 6a shows the fabricated Vernier filter comprising two rings of different sizes, where light is coupled into the add port of the larger ring and extracted from the drop port of the smaller ring. Resonance peaks ideally appear when the resonances of the two rings are aligned.[31] Figure 6b shows the transmission spectra of the fabricated Vernier filter, where a ~0.2 nm mismatch between the resonance wavelengths of the two rings suppresses the transmission in the highlighted region. Two implantations were applied to the larger ring (Figure 6a). The first implantation at the coupling region compensates for undercoupling, resulting in a 5 dB improvement in ER. The second implantation on the ring shifts its resonance, improving the spectral alignment between the two rings and yielding a further 10 dB increase in ER.

## 2.4 | Insertion loss and long-term stability

Minimizing trimming-induced excess loss is essential to avoid loss accumulation in large-scale PICs. We evaluate the excess loss of silicon ion implantation using microring resonators at a dose of 1.55 nC/μm$^2$, consistent with that used for DC and microring tuning in this work. The calculated excess loss is less than 0.001 dB/π, owing to the negligible absorption of silicon at 1550 nm and the confinement of the Si-implanted region to the cladding, thereby minimizing additional scattering loss.

Representative non-volatile post-fabrication trimming approaches are compared in Table 1. Considerable efforts have been devoted to reducing the excess loss associated with post-fabrication trimming through different physical mechanisms. These approaches achieve low excess loss and are suited to various application scenarios. The proposed silicon ion implantation approach achieves similarly low excess loss while enabling single-step localized trimming of cladded PICs.

TABLE 1 | Comparison of state-of-the-art non-volatile post-fabrication trimming approaches

| Technique | Platform | λ (μm) | Cladding | Loss (dB/π) | Temp. (°C) | Ref. |
|---|---|---|---|---|---|---|
| Si ion implantation | $Si_3N_4$ | 1.55 | Yes | <0.001 | RT | This work |
| $Sb_2S_3$ | Si | 1.55 | Yes | 0.72 | ~300 | [32] |
| $Sb_2Se_3$ | $Si_3N_4$ | 1.55 | No | 1.15 | ~190 | [33] |
| UV laser trimming | $Si_3N_4$ | 1.95 | Yes | ~0.23 | RT | [28] |
| Ge implantation & anneal | Si | 1.55 | No | ~0.2 | ~450 | [27] |
| FIB carbon deposition | $Si_3N_4$ | 1.55 | No | 0.35 | RT | [34] |

Long-term stability is essential for the practical deployment of non-volatile post-fabrication trimming. We assess the proposed approach through repeated measurements over a 120-day period, with the chips stored under ambient conditions and no post-treatment applied after implantation. Figure 7a shows the mean coupling ratio and its variation for five identical DCs after implantation, confirming stable device performance over the monitoring period. Figure 7b shows that the resonance shift of microring resonators remains stable over 25 days across implantation doses of 0.001 to 5 nC/μm$^2$.

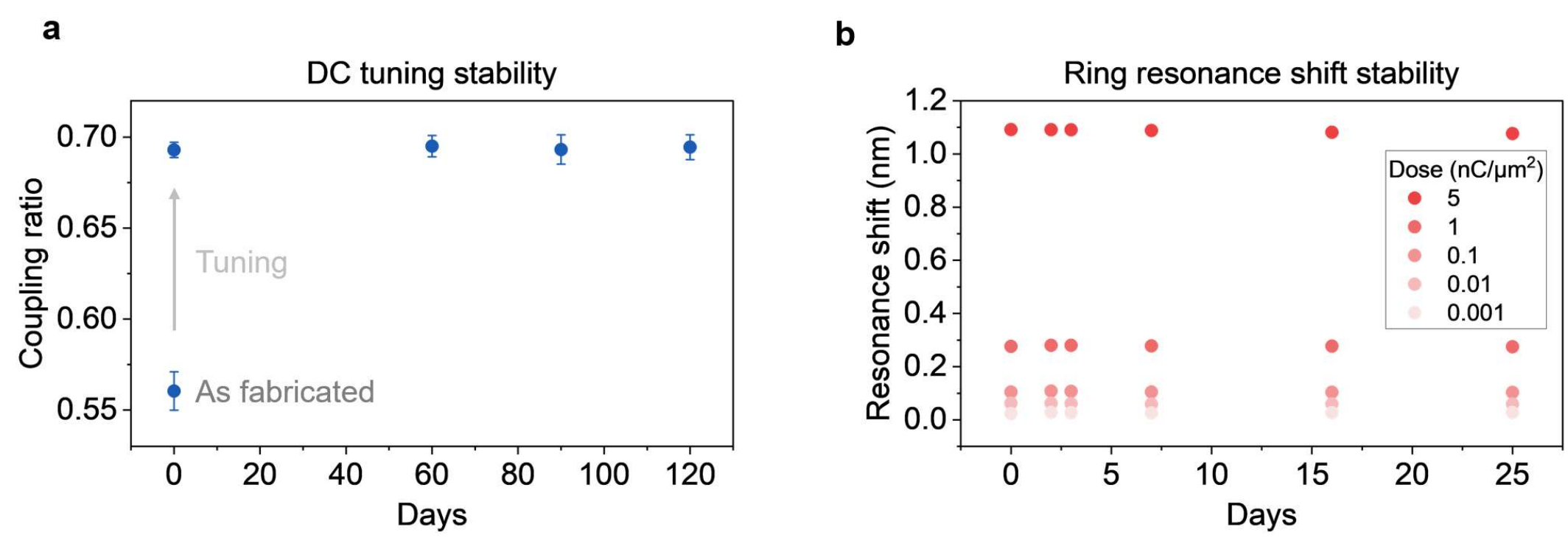

**FIGURE 7 |** Device response stability after trimming. (a) CR of trimmed DCs measured over 120 days. (b) Resonance shift of microring resonators trimmed with different FIB doses, measured over 25 days.

## 3 | Conclusion

We have demonstrated a non-volatile, ultra-low-loss, and localized post-fabrication trimming approach for cladded PICs based on silicon ion implantation. By confining the Si-implanted region to the cladding rather than directly modifying the waveguide, the approach achieves localized index modification with an excess loss below 0.001 dB/π. We show that precise response tuning of representative photonic building blocks, including DCs and microring resonators, can be realized by controlling the implantation dose, geometry, and position. The response after trimming remains stable over a four-month observation period under ambient conditions. Automated trimming was demonstrated on a PCA, reducing the average output variation from 78.5% to 5.9% and thereby demonstrating the scalability of the approach from individual devices to photonic circuits. The approach was further validated on the $LiNbO_3$ platform, confirming its cross-platform applicability. As PICs become increasingly complex, mitigating the accumulation of fabrication-induced response variations becomes essential to ensure reliable circuit operation. Together, these capabilities establish a practical route toward precise post-fabrication trimming of cladded PICs, facilitating the deployment of larger-scale PICs.

## 4 | Method

### 4.1 | Device fabrication

$Si_3N_4$ chips consist of a 335 nm $Si_3N_4$ layer and a 3300 nm $SiO_2$ layer on a Si substrate. The on-chip photonic circuits were fabricated using electron-beam lithography (EBL, 100 kV Raith EBPG 5150), followed by reactive ion etching (RIE, Oxford PlasmaPro 80). The $Si_3N_4$ waveguides were fully etched. An HSQ solution diluted in MIBK was then spin-coated, followed by EBL exposure and development to form a nominally 500 nm-thick HSQ top cladding.

Thin-film $LiNbO_3$ chips consist of a 300 nm $LiNbO_3$ layer and a 4700 nm $SiO_2$ layer on a Si substrate. The photonic circuits were fabricated using the same EBL process, followed by inductively coupled plasma etching (ICP, Oxford PlasmaPro 100). The $LiNbO_3$ waveguides were partially etched to a depth of 275 nm. HSQ cladding was subsequently formed using the same spin-coating and EBL process.

### 4.2 | Device tuning with silicon ion implantation

A Raith VELION FIB was used for silicon ion implantation with a 35 kV Si++ beam and a 30 μm aperture. Using pre-defined gold alignment markers and pre-loaded design files, automated alignment enabled sequential implantation of multiple PIC devices.

### 4.3 | Photonic characterization

Transmission spectra before and after tuning were measured using a tunable laser (Santec TSL-110), a polarization controller, and photodetectors (Newport 2011-FC). The chip was mounted onto an x-y translation stage, with light coupling achieved through a fiber array positioned above the chip. The laser and photodetector were computer-controlled for wavelength sweeping and data acquisition. Unless otherwise stated, error bars represent the standard deviation across three or more nominally identical devices.

### 4.4 | Calculation of excess loss (dB/π)

The excess loss per π phase shift (dB/π) is used as a figure of merit to evaluate the optical loss of post-fabrication trimming techniques. It represents the excess loss required to induce a π phase shift, with lower values indicating a smaller penalty to device performance. For ring resonators, the induced phase shift is determined from the measured resonance wavelength shift normalized by the free spectral range (FSR), where one FSR corresponds to a phase shift of 2π. The introduced loss is extracted based on the difference between the propagation losses calculated before and after trimming, together with the device geometry. The excess loss per π phase shift is then obtained as the introduced loss divided by the corresponding induced phase shift.

### 4.5 | Materials characterization

Electron-transparent lamellae were prepared from the $Si_3N_4$ chip with the FIB technique using a ZEISS Crossbeam 340 SEM/FIB system. The lamellae were then characterized in STEM mode with a Titan Themis G3 300 operated at 300 kV at an extraction voltage of 3.45 kV for the field emission gun (X-FEG). Elemental mapping was performed with a quadruple EDS detector. EELS spectrum imaging was carried out in STEM mode at 300 kV using a Quantum 965 ER Gatan imaging filter (GIF).

**Acknowledgements**

R.X. gratefully acknowledges the Alexander von Humboldt Foundation for providing a postdoctoral fellowship. W.H.P.P. acknowledges support from the Deutsche

Forschungsgemeinschaft (DFG, German Research Foundation) under project numbers 390761711 and 390900948. This work was also supported by the European Union's Horizon projects (HYBRAIN, grant no. 101046878; 2DNEURALVISION, grant no. 101119489) and the European Research Council (PICNIC, grant no. 101200429) to W.H.P.P.

**Data Availability Statement**

The data that support the findings of this study are available from the corresponding author upon reasonable request.